\documentstyle[epsf,seceq]{ptptex}

\title{
Dileptons and Photons at the CERN SPS
}

\author{
Itzhak {\sc Tserruya}\footnote{E-mail address: tserruya@ceres.weizmann.ac.il}}

\inst{
    Department of Particle Physics, Weizmann Institute of Science \\ 
                         Rehovot 76100}

\abst{

CERES, HELIOS-3 and NA38 have observed a significant excess of lepton
pairs in collisions of 200 GeV/nucleon S beams with heavy targets in comparison to the 
expected yield from
known hadronic sources. The excess was confirmed by  results obtained by  CERES 
and  NA50 experiments with the Pb beam of 160 GeV/nucleon.
The enhancement is more pronounced in the low-mass region
(m = 0.2 - 1.5 GeV/c$^2$) measured close to mid-rapidity and seems to hint at 
in-medium modifications of the vector mesons and in particular at a decrease of the 
$\rho$-meson mass interpreted as a precursor of chiral symmetry restoration. 
On the other hand, there is no evidence of a signal in the search for direct photons  
and the various measurements, carried out by the CERES, HELIOS-2 and WA80 experiments,
allow to set an upper limit of the order of 10\%.
This paper reviews all experimental results on dileptons (electron and muon pairs)
with invariant mass m $<$ 3 GeV/c$^2$ and real photons obtained from the CERN 
SPS heavy-ion program with S and Pb beams,
together with  the current attempts to understand them.} 

\begin{document}

\maketitle

\section{Introduction}

   At the 1995  Quark Matter Conference, the three experiments involved 
in the measurement of dileptons at the CERN SPS --CERES, HELIOS-3 and NA38-- presented 
evidence of a large  excess  of dileptons --in particular at low-masses
(m = 0.2 -- 1.0 GeV/c$^2$)  but also at intermediate masses (m = 1.5 -- 2.5 GeV/c$^2$)--
in S-induced collisions at 200 GeV/nucleon \cite{it-qm95}. Since then, these results 
have been at the focus of attention, triggering a strong interest 
mainly stimulated by the possibility that the low-mass excess could result from the 
decrease of the $\rho$-meson mass as a precursor of chiral symmetry restoration 
\cite{li-ko-brown}.  
Recent results obtained with the Pb beam by the CERES and NA50 experiments show also 
an enhanced production of dileptons at low and intermediate masses confirming at least
qualitatively the results with the S beam \cite{tu-qm96,pb-au-preprint,scomparin-qm96}.
  This paper gives an overview of all experimental results on dilepton continuum and 
photons obtained since the beginning of  the CERN SPS heavy-ion program and of the current 
attempts to understand them.

The main goal of ultra-relativistic heavy-ion collisions is the study of hadronic matter
under extreme conditions of density and temperature and in particular to search for
evidence of the predicted phase transition(s) \cite{tdlee} leading to Quark Gluon 
Plasma formation --where quarks and gluons are free to move over a large volume compared
to the typical size of hadrons-- and to chiral symmetry restoration where masses drop
to zero. The importance of dileptons --$e^+e^-$ or $\mu^+\mu^-$ pairs --, 
in this endeavour 
has been emphasized time 
and again since it was first proposed by Shuryak in 1978 \cite{shuryak78}. 
Since  dileptons interact only electromagnetically, 
their mean free path is relatively large compared to the size of the system formed 
in these collisions;
therefore, once produced they can leave the interaction region 
and reach the detectors without any further interaction, carrying information about 
the conditions and properties of the matter at the time of their production
and in particular of the early stages when temperature and energy density have 
their largest values. This has to be contrasted with the hadronic observables 
which are sensitive to the late stages of the collisions at and after freeze-out, 
i.e. once the hadronic system stops interacting.
 
  The main topic of interest is the identification of 
dileptons emitted as thermal radiation which can 
tell us about the nature of the matter formed, the conjectured quark-gluon plasma 
(QGP) or a high-density hadron gas (HG). The elementary processes involved are
{\it q}{\it $\overline{q}$\/} annihilation in the QGP phase and $\pi^+\pi^-$ 
annihilation in the HG phase. Since 
the thermal emission rate is a strongly increasing function of the temperature, it is  
most abundantly produced at the early 
stages when the temperature and energy density have their largest values, 
thereby providing a higher sensitivity to identify the thermal radiaton from the 
QGP, if it is formed. This sensitivity increases as the initial temperature of the
system increases relatively to the critical temperature of the phase transition. 
Theoretical calculations have singled out the mass range of 1-3 GeV/c$^2$ as the most 
appropriate window to observe the thermal radiation from the QGP 
phase~\cite{kajantie86,ruuskanen92} at initial temperatures likely to be reached at 
RHIC or LHC. At SPS energies, the initial temperature is believed to be
close to or below the critical temperature, and therefore one expects
the dense hadron 
gas to be the dominant source of thermal radiation. The window to search for 
it is at low masses, around and below the $\rho$-meson 
mass \cite{kajantie86,cleymans91,koch93}, since the $\pi^+\pi^-$ annihilation cross 
section is dominated by the pole of the pion electromagnetic form 
factor at the $\rho$ mass.

The physics potential of dileptons is further emphasized by the 
capability to measure the  vector mesons --which are considered as important
messengers of the collision dynamics \cite{shor,heinz}--
through their leptonic decays. Of particular interest is 
the decay of the $\rho$ meson into a lepton pair since it provides a unique experimental
window to observe the effects of chiral symmetry restoration. Due to its very short
lifetime ($\tau$ = 1.3 fm/c) compared to the typical fireball lifetime of 10-20 fm/c
at SPS energies, most of the $\rho$ mesons produced in the collision will decay
inside the interaction region with a reduced or even zero mass
if the temperature and or the baryon  density are large enough for chiral symmetry 
restoration to take place. The situation is very different for the other mesons, 
$\omega,\phi$ or $J/\psi$ because of their much longer lifetimes; they will be 
reabsorbed in the medium or they will decay well outside the interaction region where 
they have regained their vacuum masses.
   
 Together with this brief discussion on the physics interest and physics potential of 
dileptons,  
 one should also appreciate the difficulties associated with the measurements.   
The main experimental problem is the huge {\it combinatorial background} of 
uncorrelated lepton pairs originating from the decay of hadronic particles 
(and also from conversions in the measurement of electron pairs), which strongly 
increases as the coverage moves to the low-mass and low-p$_t$ regions. 
Secondly, dileptons can be emitted by a variety of sources and therefore,
before one can make a claim on the observation of any new effect, 
it is absolutely necessary to have   
a thorough understanding of the contribution from all known sources, i.e. the
{\it physics background}.  Drell-Yan and semi-leptonic decays of
charm mesons, produced in the  primary hard collisions, are the main 
contributions at intermediate masses. In the low-mass region, the physics 
background is dominated by the electromagnetic decay of hadrons
(Dalitz decays of $\pi^o$,$\eta$,$\eta$' $\rightarrow$ $l^+l^-\gamma$, 
$\omega$ $\rightarrow$  $l^+l^-\pi^o$,
and resonance decays of $\rho,\omega,\phi$ $\rightarrow$  $l^+l^-$)
which mostly take  place at a late stage of the collision, long after freeze-out.
To understand the physics background, the experiments have adopted a systematic 
approach, performing precise measurements of the dilepton production
in pp and pA collisions. These studies provide  also the basis for comparison 
to nucleus-nucleus collisions and to identify any possible deviation from
known physics. The  results which I will review below show clear evidences
of such deviations.

 The above discussion is also relevant to real photons, since real and 
virtual (dileptons) photons are expected to carry 
the same physics information. However, the physics background for real photons 
is larger by orders 
of magnitude as compared to dileptons, making the measurement of photons much 
less sensitive to a new source.
 
This paper is organized as follows. Section 2 briefly presents and lists all
measurements of electromagnetic observables performed so far. The experimental results
are presented in Section 3. Section 4 reviews the current theoretical attempts to
understand the various results and Section 5 contains a short summary and
discussion on open  questions and further work.

\begin{table}[h!]
\caption{ List of Dilepton Measurements}
\label{tab:dileptons}
\vspace{0.2cm}
\begin{center}
\leavevmode
\begin{tabular}{|c|c| c| c|  c  |}\hline

 Experiment &  Probe      &  System           &  $y$      &  Mass (GeV/c$^2$)\\\hline
            &             & p-Be,Au 450 GeV/c &           &                  \\ 
 CERES      &   $e^+e^-$  & S-Au    200 GeV/u &  2.1-2.65 & 0 -- 2.0         \\
            &             & Pb-Au   158 GeV/c &           &                  \\\hline
 HELIOS-3   &$\mu^+\mu^- $& p-W,S-W 200 GeV/u & $>3.5$    & 0.3 -- 4.0       \\\hline
 NA38       & $\mu^+\mu^-$& p-A,S-U 200 GeV/u & 3.0-4.0   & 0.3 -- 6.        \\
 NA50       &             & Pb-Pb   158 GeV/u &           & 0.3 -- 7.0       \\\hline

\end{tabular}
\end{center}
\end{table}

\section{Measurements of electromagnetic probes}
  
   The three experiments involved in the measurement of dileptons are listed in 
Table 1. CERES is the only experiment dedicated to the 
measurement of low-mass electron pairs, from $\sim$ 50 MeV/c$^2$ up to 
$\sim $2 GeV/c$^2$, limited at the upper end by the available statistics. It covers
the mid-rapidity region with a very broad range of $p_t$. The other two experiments 
are dedicated to the measurement of muon pairs. HELIOS-3  measured a very 
broad mass range from the dimuon threshold up to the $J/\psi$ covering
mostly the forward rapidity region. NA38 (and its successor NA50)  is 
mostly focussed in 
the mass range around and below the $J/\psi$ and covers the mid-rapidity 
region. The reference measurements on pp and pA performed by  each experiment
are also listed in Table 1. HELIOS-3 has finished data-taking
whereas CERES and NA50 are continuing their physics programme with the Pb beam.

Two experiments are presently involved in the measurement of real photons
(see Table~2),
WA80 which measures them directly and CERES which uses the conversion
method. Both experiments cover almost the same mid-rapidity interval. For
completeness, Table~2 also includes HELIOS-2 which has performed the
first search of direct photons using the O and S beams of the CERN SPS.

\begin{table}[h!]
\caption{List of Real Photon Measurements}
\label{tab:photons}
\vspace{0.2cm}
\begin{center}
\leavevmode
\begin{tabular}{|c|c| c| c |}\hline

 Experiment    &  System           &   $y$      &  $p_t$ (GeV/c)    \\\hline
CERES          &  S-Au 200 GeV/u   &   2.1-2.7  &    0.4-2.0        \\\hline 
HELIOS-2       & p,O,S-W 200 GeV/u &   1.0-1.9  &    0.1-1.5        \\\hline
WA80           &  O-Au   200GeV/u  &   1.5-2.1  &    0.4-2.8        \\

               &  S-Au   200 GeV/u &   2.1-2.9  &    0.5-2.5        \\\hline

\end{tabular}
\end{center}
\end{table} 
\vspace{-0.6cm}

\section{Experimental Results}

   The three dilepton experiments have  reported an excess of dileptons in S-induced
interactions over the known sources as measured in pp or pA collisions after scaling  
them to the S-nucleus case. This excess is confirmed, at least qualitatively, by the
results obtained with the lead beam.
\vspace{-0.5cm}
\begin{figure}[h!]
\begin{center}
\leavevmode
\epsfxsize=7cm
\epsfysize=8cm
\epsffile{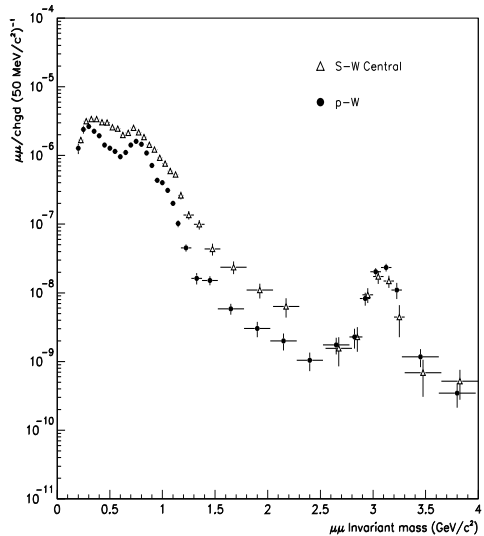}
\end{center}  
\vspace{-0.5cm}
\caption{Dimuon invariant mass spectra measured by HELIOS-3
in p-W and in central S-W interactions  at 200 GeV/nucleon.
\protect\cite{masera-qm95}. }
\label{fig:helios3}
\end{figure}

 The overall picture is beautifully illustrated in Fig. 1 which shows the
results of HELIOS-3 in  central S-W interactions together with those
obtained in p-W  at 200 GeV/nucleon \cite{masera-qm95}.
They are presented in the form of muon pairs per charged particle 
measured in the same rapidity interval. The enhancement 
covers a very broad mass range including
the low-mass continuum (m = 200 - 600 MeV/c$^2$), the vector
mesons $\rho$, $\omega$ and $\phi$, and the intermediate-mass continuum
(above the $\phi$ and below the $J/\psi$ masses). The figure also
displays the well known $J/\psi$ suppression, a topic of much current interest
 (see the paper presented by D. Kharzeev at this school
 \cite{kharzeev}). In the following we discuss in detail the experimental 
results on low- and intermediate-mass pairs.

\subsection{Low-mass Dileptons}

\subsubsection{Results with the p and S beams}
 
 The low-mass region is systematically studied by the CERES experiment.  
Fig.~2  shows  the low-mass spectra measured in
450 GeV/c p-Be (a very good approximation to the p-p system) and p-Au 
\begin{figure}[h!]
\begin{center}
\leavevmode
\epsfxsize=7.8cm
\epsfysize=6cm
\epsffile{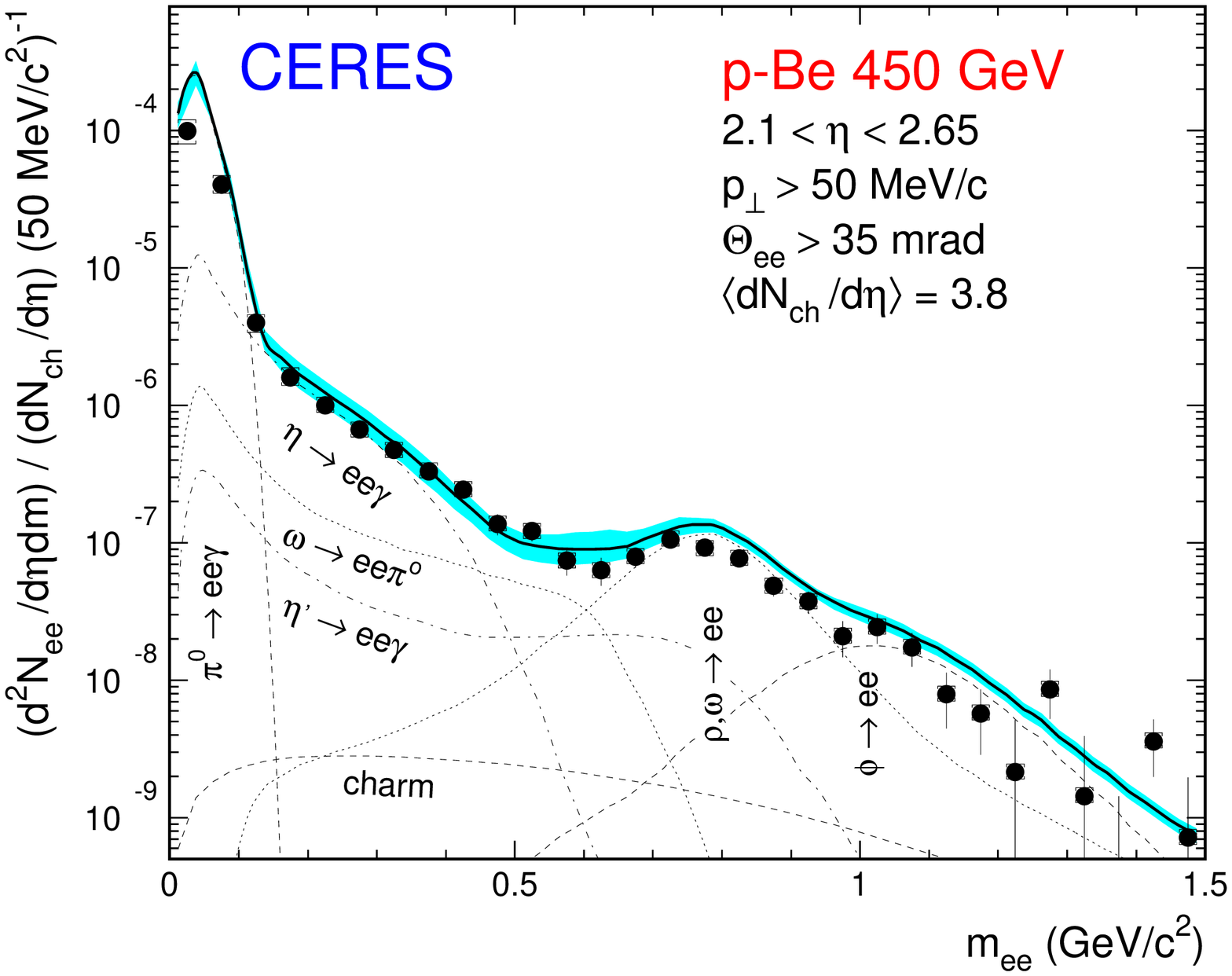}
\epsfxsize=7.8cm
\epsfysize=6cm
\epsffile{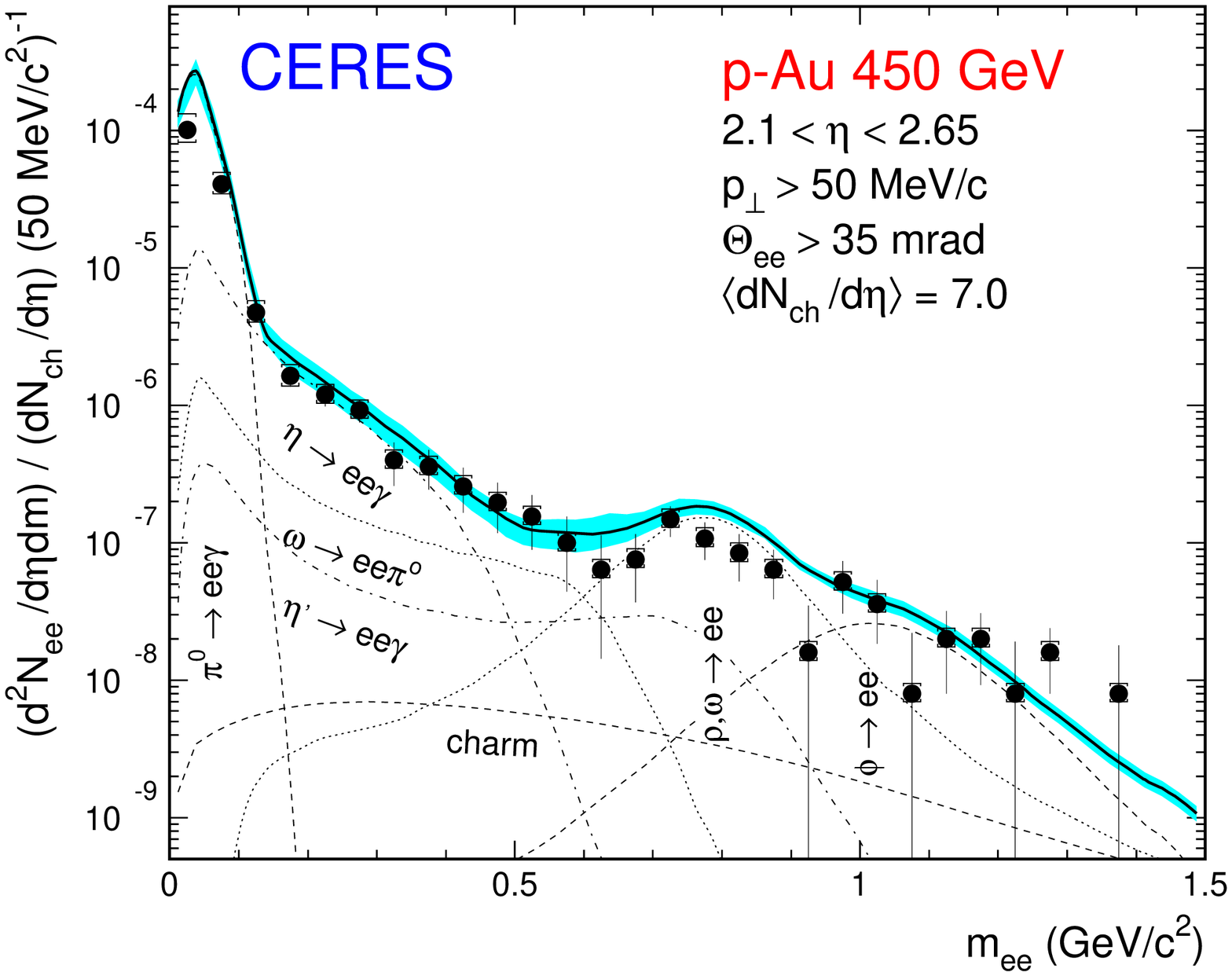}
\end{center}  
\caption{$e^+e^-$ invariant mass spectra in 450 GeV/c p-Be and p-Au collisions measured
by CERES \protect\cite{p-paper}. Statistical (vertical bars) and systematic (brackets)
errors are plotted independently of each other. See text for details.}
\end{figure}
collisions \cite{p-paper}.  The data are normalized 
to give the pair density per charged particle density within the acceptance 
of the CERES spectrometer. The lines represent the contributions from the
known hadron decays and the shaded area gives  the systematic error
on the summed contributions. It is assumed that the particle production ratios of
these sources is independent of the collision system and  
consequently, that the $e^+e^-$ production scales with the number of charged 
particles. One sees that the p-induced data are very  well reproduced by  
electron pairs from the known hadronic sources. There is no need to invoke 
any unconventional or 
"anomalous" source of lepton pairs (see also ref. 17). The situation is 
completely
different in the nucleus-nucleus case.	
The measured S-Au mass spectrum \cite{prl95}, shown in Fig. 3, reveals a dramatic 
effect; it has a different shape and shows a strong enhancement 
over the hadronic contributions at masses m ~$>$ ~0.2 ~GeV/c$^2$,
reaching even one order of magnitude around m $\sim$ 0.4 GeV/c$^2$.
The enhancement factor --defined as the ratio of the measured  yield
integrated over the mass range m = 0.2 - 1.5 GeV/c$^2$ to the expected 
one-- is   5.0 $\pm$ 0.7 (stat.) $\pm$ 2.0 (syst.) 

\begin{figure}[h!]
\begin{center}
\leavevmode
\epsfxsize=10cm
\epsfysize=9cm
\hspace{-1.2cm}
\epsffile{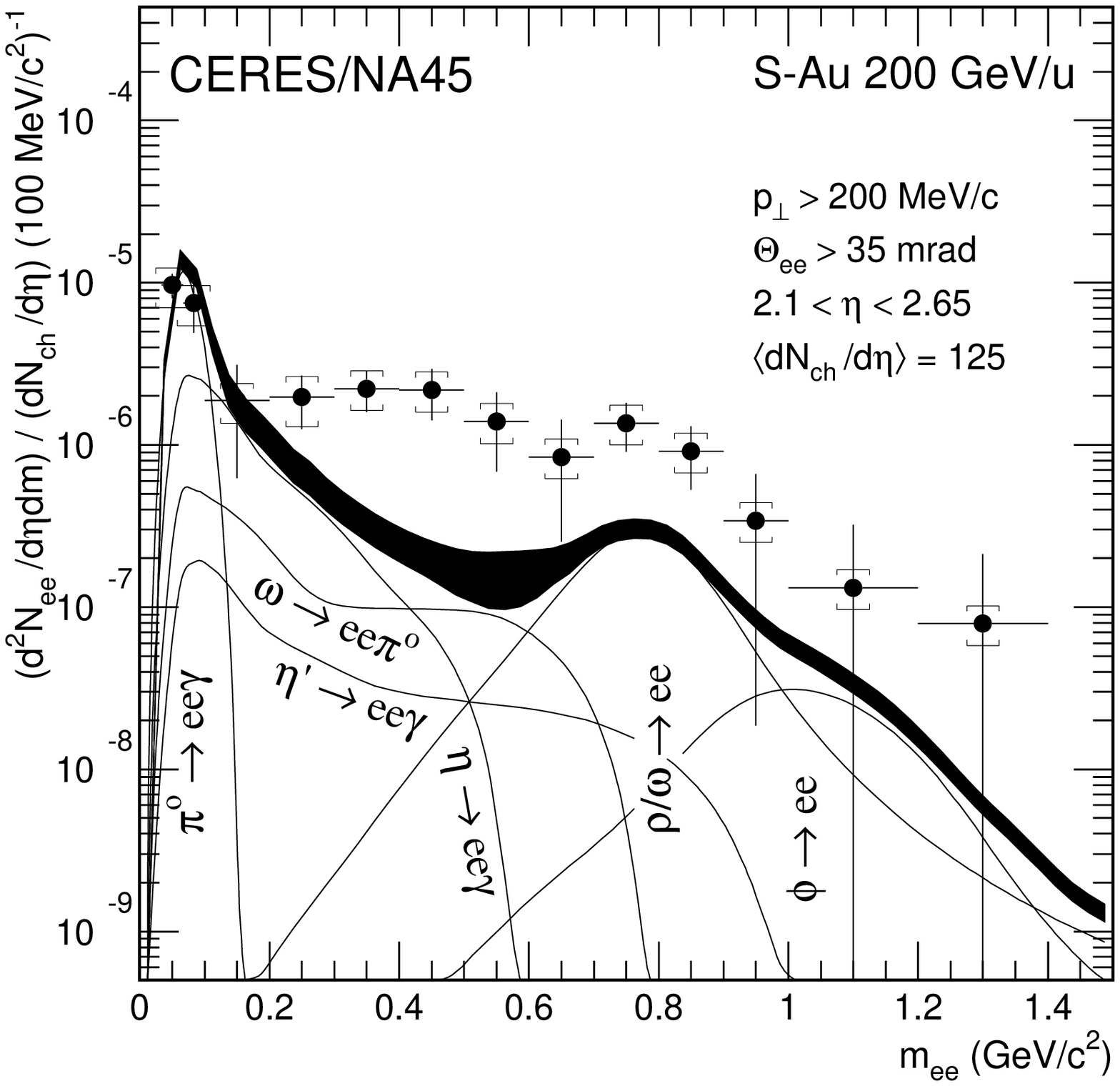}
\end{center}  
\caption{$e^+e^-$ invariant mass spectra in 200 GeV/nucleon S-Au collisions measured
by CERES \protect\cite{prl95}. Statistical (vertical bars) and systematic (brackets)
errors are plotted independently of each other. See text for details.}
\end{figure} 

The reliability of the results presented in Fig. 3 is affected by  the limited 
statistics of the sample,  (a total signal, for masses m $>$
200 MeV/c$^2$, of 445 pairs with a signal to combinatorial background ratio 
of 1/4.3). Furthermore,  the shape of the excess and the combinatorial
background are quite similar for masses m $>$ 200 MeV/c$^2$
raising the question whether the combinatorial background  has been 
subtracted  correctly \footnote{The amount of combinatorial background in the $e^+e^-$
measured yield is assumed to be equal to the total like-sign yield, $e^+e^+ + e^-e^-$,
such that the signal is obtained by subtracting the like-sign yield from the
unlike-sign yield}. However, it seems that the similar shapes is a mere coincidence.
The results remain stable within errors
all along the chain of rejection cuts and when the cut values are varied by 
$\sim \pm$15\% around the optimal values \cite{pb-au-preprint}.

\subsubsection{Results with the Pb beam}

Figure 4 shows the recent results of CERES obtained in Pb-Au collisions at
160 GeV/nucleon \cite{tu-qm96,pb-au-preprint} with an average charge multiplicity of 
$<$dn/dy$>$ = 220 corresponding to the top $\sim$ 35 \% of the geometrical
cross section. The spectrum is based on a total of 650 pairs with a signal
to background ratio of 1/8. The results shown in Fig. 4 are very similar to 
those obtained in S-Au collisions. The yield is clearly enhanced compared to the 
predicted one from hadron decays. This is most pronounced in the region from 
300 to 700 MeV/c$^2$ where the 
enhancement factor is 5.8 $\pm$ 0.8 (stat) $\pm$ 1.5 (syst). The enhancement 
extends also to higher masses. In the larger mass interval
0.2 $\le$ m $\le$ 2.0 GeV/c$^2$ the enhancement factor
 is here 3.5 $\pm$ 0.4 (stat) $\pm$ 0.9 (syst).
\begin{figure}[h!]
\begin{center}
\leavevmode
\epsfxsize=7cm
\epsfysize=8cm
\epsffile{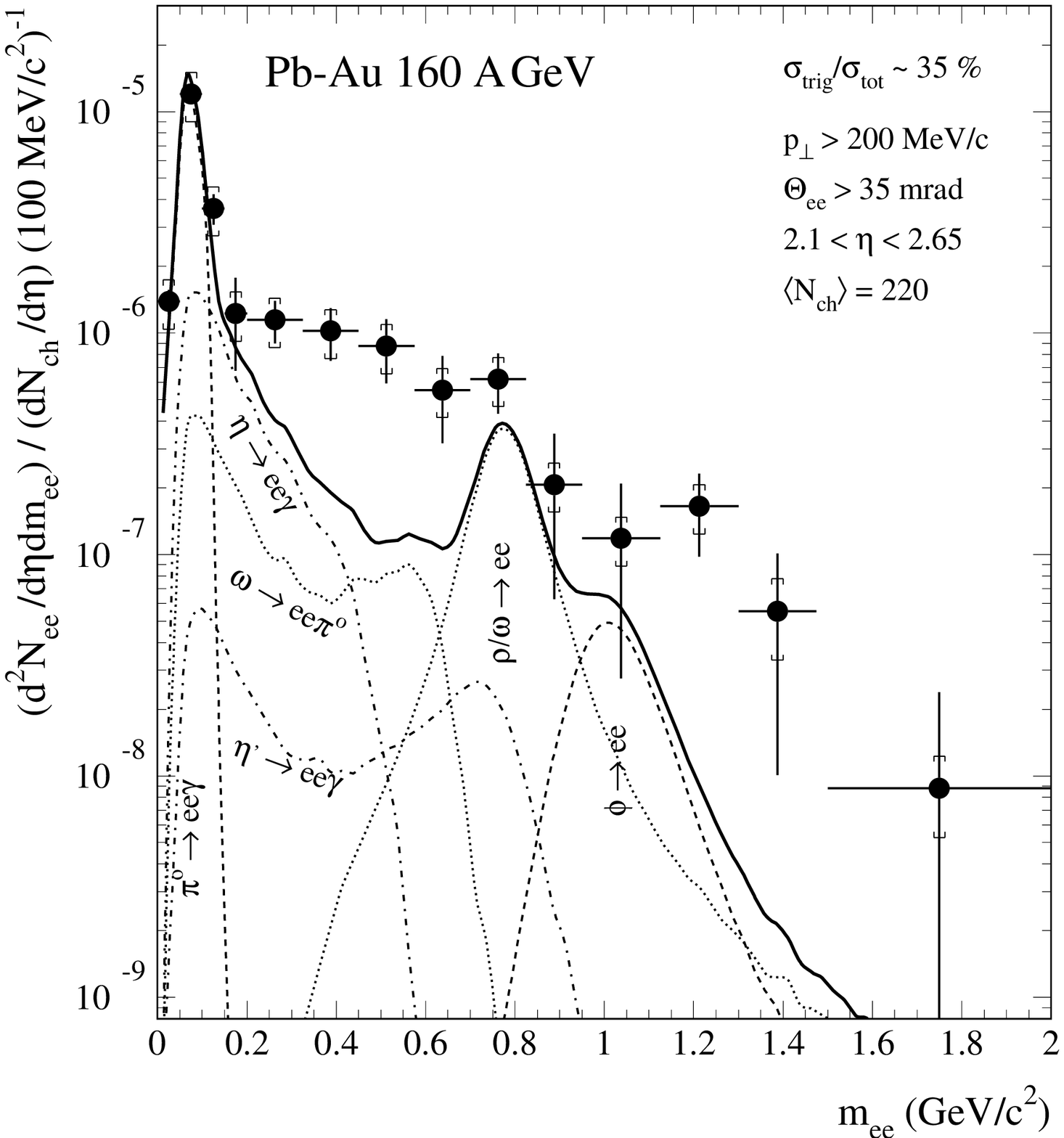}
\end{center}  
\vspace{0.2cm}
\caption{$e^+e^-$ mass spectrum measured by CERES in 
160 GeV/nucleon Pb-Au collisions ~\protect\cite{pb-au-preprint}.}
\end{figure}

   An important aspect to characterize the excess  is provided by 
the multiplicity dependence of the dilepton yield. The thermal radiation
emitted either by partons or by pions should exhibit a quadratic dependence with 
multiplicity for a fixed interaction volume, since the emission rate is 
proportional to the product of the particle and anti-particle densities.
In spite of the limited statistics of the Pb sample an attempt was made to exploit 
the large range of impact parameters covered by the data to study the yield as a function
of the event multiplicity.
\begin{figure}[h!]
\begin{center}
\leavevmode
\epsfxsize=7cm
\epsfysize=8cm
\epsffile{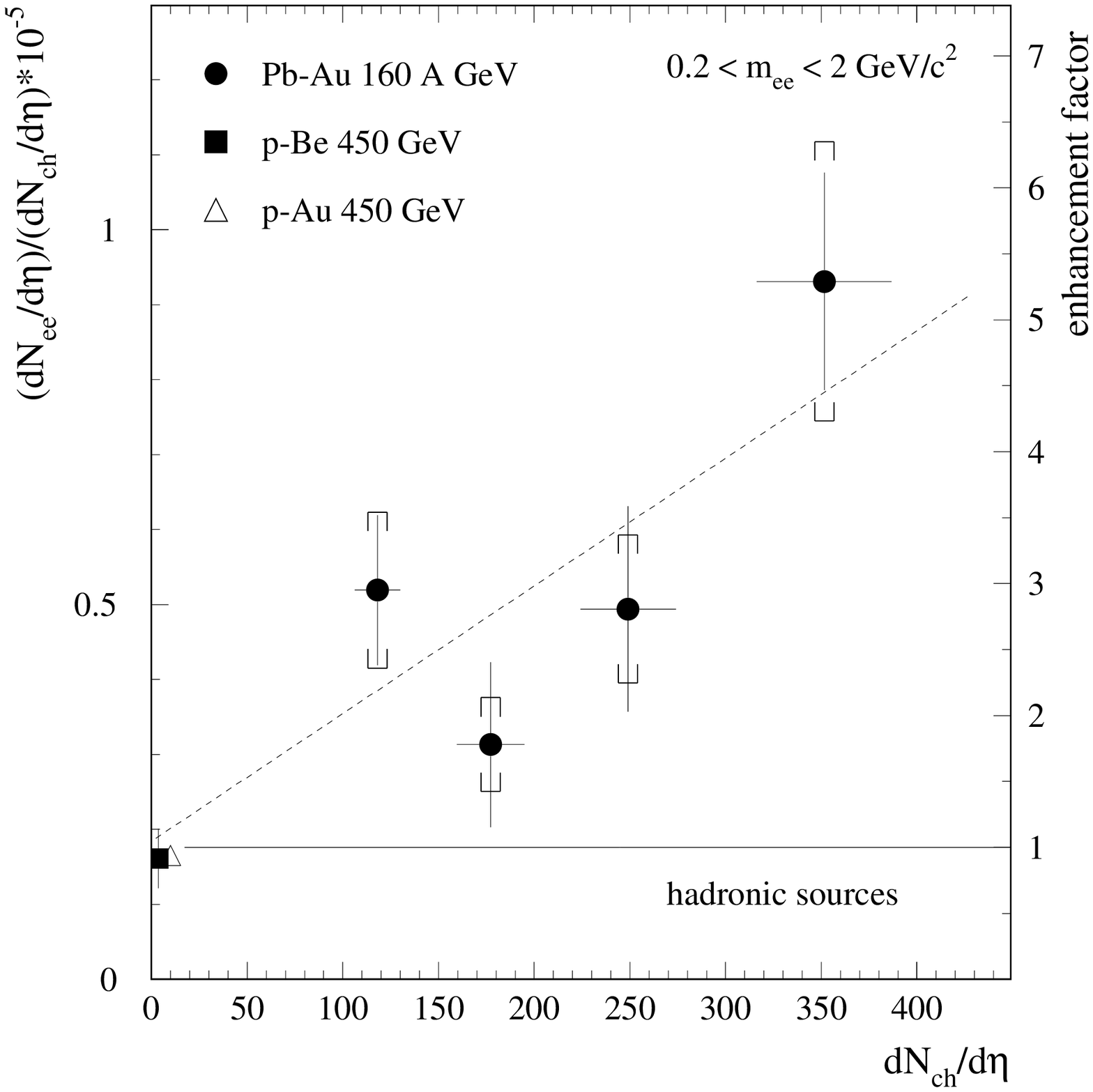}
\end{center}  
\vspace{0.2cm}
\caption{Multiplicity dependence of the $e^+e^-$ pair yield normalized to the charged
particle density measured by CERES in Pb-Au collisions at 160 GeV/nucleon. The horizontal
line indicates the expectation from hadron decays \protect\cite{pb-au-preprint}.}
\end{figure} 
 Fig. 5 shows the pair density per charged particle density
integrated over the mass range m $>$ 200 MeV/c$^2$ for four bins of multiplicity. 
If hadron decays were to be the only source of electron pairs --as it is the case 
in pp and pA collisions--  the data should scale linearly with multiplicity. Therefore 
the pair density normalized to the charged-particle density should remain constant 
at the level determined by  pp collisions and shown by the horizontal line in the figure. 
Although the error bars are large, one observes a clear  deviation from the constant 
hadronic level indicating that the excess increases faster than linearly with 
multiplicity. This result confirms previous hints of a non-linear scaling observed 
in the S-Au case \cite{cph-thesis}.
  
\subsection{Intermediate-mass Dileptons}

   The excess in the intermediate mass region is best illustrated by the 
results of NA38 and NA50 \cite{scomparin-qm96,ramos-qm95}.
\begin{figure}[h]
\begin{center}
\leavevmode
\epsfxsize=13cm
\epsfysize=10cm
\hspace{-1.2cm}
\epsffile{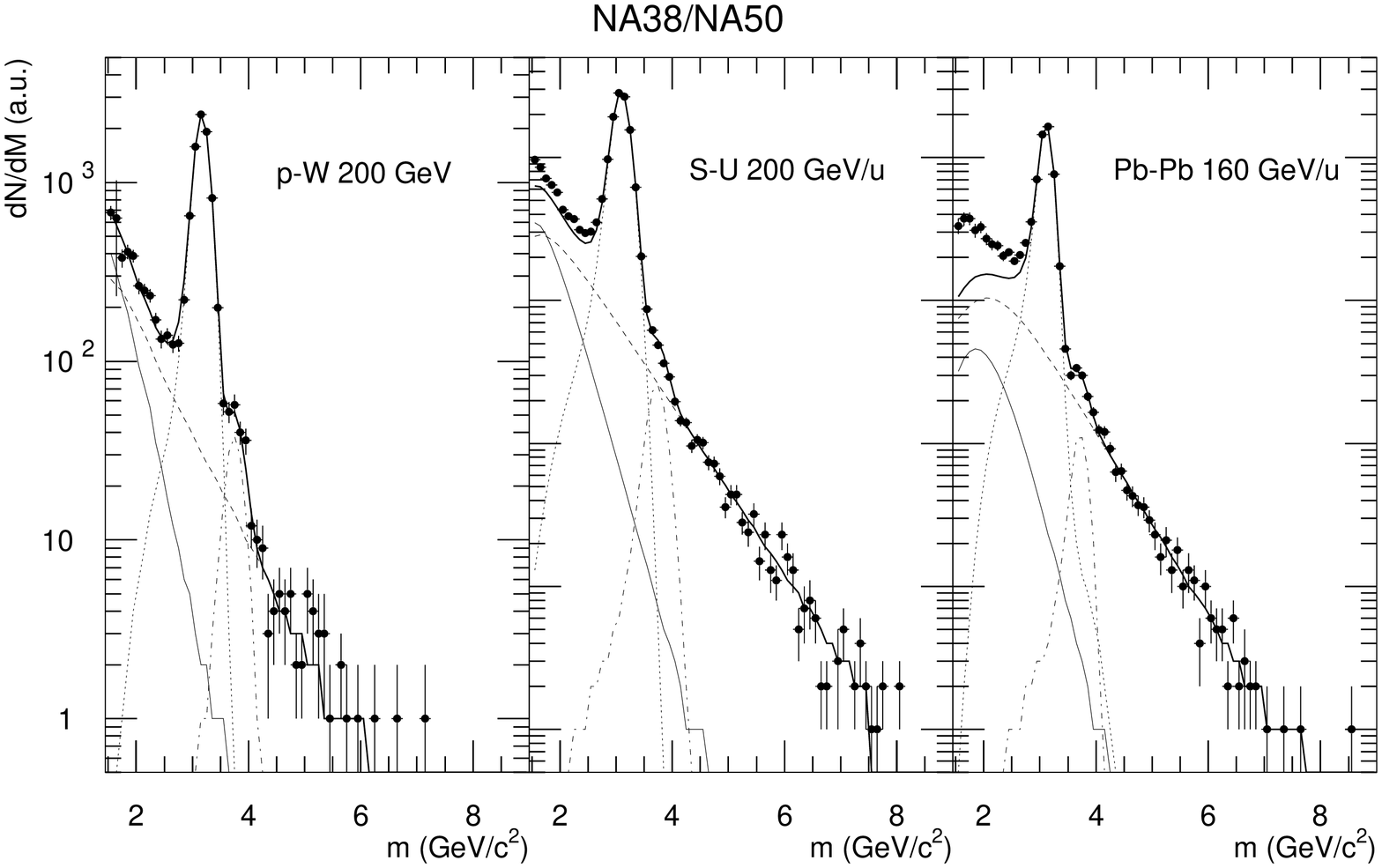}
\end{center}  
\caption{Di-muon invariant mass spectrum measured by NA38 in 200 GeV/nucleon p-W
and S-U collisions and by NA50 in 160 GeV/nucleon Pb-Pb 
collisions~\protect\cite{scomparin-qm96}. 
The dotted, dashed, dot-dashed, thin and thick lines represent the J/$\psi$,
Drell-Yan, $\psi^{'}$, open charm and total contributions, respectively.}
\end{figure}
Fig. 6 shows their results in p-W, S-U and Pb-Pb
collisions. Drell-Yan and semi-leptonic charm decay are the main 
contributions at masses 1.5 $<$ m $<$ 2.5 MeV/c$^2$, and they provide a 
good description of the p-W data. A small enhancement is observed in
S-U collisions which appears more pronounced in the Pb-Pb system.
In these two cases, the Drell-Yan and charm decay cross sections are
assumed to scale with the product A$_P$A$_T$ of projectile and target mass numbers.
For details on the normalization and extrapolation procedures see 
refs. 5 and 20.

\subsection{Photons}

  In contrast with the dilepton measurements, there is no evidence for enhancement
in the measurement of real photons. The three experiments
which have performed measurements of real photons have been able to establish
only an upper limit for the production of thermal photons, which is now of the order 
of 10 \%  of the
expected yield from hadron sources.
\begin{figure}[h!]
\begin{center}
\leavevmode
\epsfxsize=7cm
\epsfysize=8cm
\epsffile{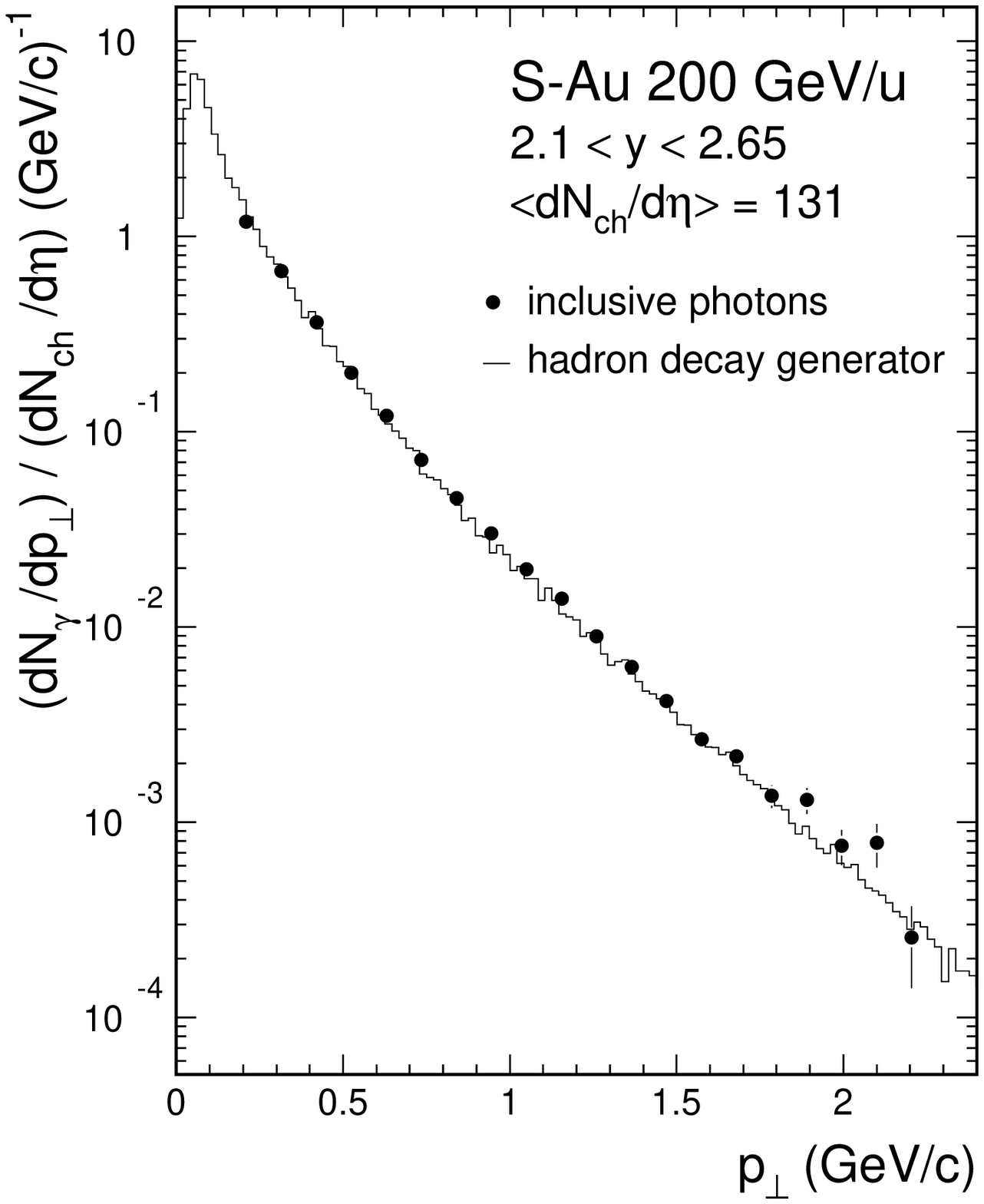}
\end{center}  
\vspace{-0.5cm}
\caption{CERES results on inclusive photon p$_t$ distribution from central S-Au 
collisions and comparison to predictions from hadron decays \protect\cite{ceres-ph}.}
\end{figure}

 The CERES and WA80 results on S-Au collisions at 200 GeV/nucleon measured at 
mid-rapidity \cite{ceres-ph,wa80-ph} are in very good agreement with each other
\cite{it-qm95} and 
the absolute yield is  very well reproduced by the expectation from
hadronic sources as illustrated in Fig. 7 with the CERES results.
From this comparison, CERES deduces a photon excess equal to 4\% of the total 
inclusive photon yield
from expected sources with systematic errors of +9\% and -14\% (the statistical
errors are negligible), namely an excess which is consistent with zero.  WA80 has 
a similar result, an excess of 5\% with smaller errors $\pm 5.8\%$.
Previous  results on this topic, published by HELIOS-2
\cite{helios2-photons} and WA80 \cite{albrecht-91} reached also the same conclusion, 
although with larger errors. In all these attempts, the sensitivity is actually limited
not by the statistical but by  the systematic errors, too large to identify 
a source which is expected to be of the order of a few percent of the total yield
(see next section).

\section{Theoretical Interpretations}

   The results presented in the previous section, and in particular the
low-mass dilepton enhancement, have triggered a strong wave of theoretical 
activity~\cite{li-ko-brown,cassing}$^{-}$\cite{wambach}. 
In this section I shall try to 
review and summarize the main highlights. 

   First of all, one may question the validity of the assumption of constant
particle production cross section ratios used to scale the expected dilepton yield 
from the pp to the ion case, although  it is certainly a very reasonable one. 
More specifically, one could  ask whether it would be possible to reproduce  
the low-mass S-Au data of CERES with a modified ``cocktail'',
using an enhanced production of the $\eta$, $\eta^{'}$ and 
$\omega$. Inspection of the S-Au spectrum of Fig. 3 reveals that this
 can  fairly well be achieved. However, this 
 would create a contradiction with other existing 
experimental information. For example, the ratio $\eta/\pi^0$ was  measured 
by WA80 in S-~S and S-Au collisions at 200 GeV/nucleon   and was 
found in very good agreement  with the  
ratio measured in pp collisions~\cite{wa80-etapi}  thus not allowing
the enhancement of the $\eta$ by the factor of 3-5 which would be  needed to bring the cocktail
close to the data. Alternatively, one would have to enhance the $\eta^{'}$ or/and the
$\omega$ yields by at least one order of magnitude; such a dramatic increase 
would have observable consequences in the real photon yield 
(since these particles have a large branching ratio to decay into photons) creating a 
conflict with the data of CERES and WA80 discussed previously. We therefore
conclude  that it is not possible to explain the observed low-mass dilepton
enhancement by a modification of the particle ratios in the cocktail of 
known hadronic sources  (see ref. 37 for upper limits on the $\eta$ and 
$\eta^{'}$ production).  
 
   The explanation  of the low-mass excess requires therefore an additional source,
not present in the pp case. The two-pion annihilation channel  
$\pi^+\pi^- \rightarrow e^+e^- $ which is expected to become important in a high 
density environment, is an obvious candidate. The characteristic features of the excess 
--its onset at a mass ${m}_{ee}\sim 2{m}_\pi$, its extension to the low-mass 
region below and around the $\rho$-meson, and the possibility of a quadratic 
dependence with multiplicity \cite{tu-qm96,pb-au-preprint,prl95,cph-thesis}-- suggest 
indeed  
that the excess is due to this channel. This would then be the first indication 
of thermal radiation emitted from the dense hadronic matter formed in relativistic 
heavy ion collisions.

 Numerous calculations have been reported, which have all included the pion 
annihilation channel.
\begin{figure}[h!]
\begin{center}
\leavevmode
\epsfxsize=7cm
\epsfysize=8cm
\epsffile{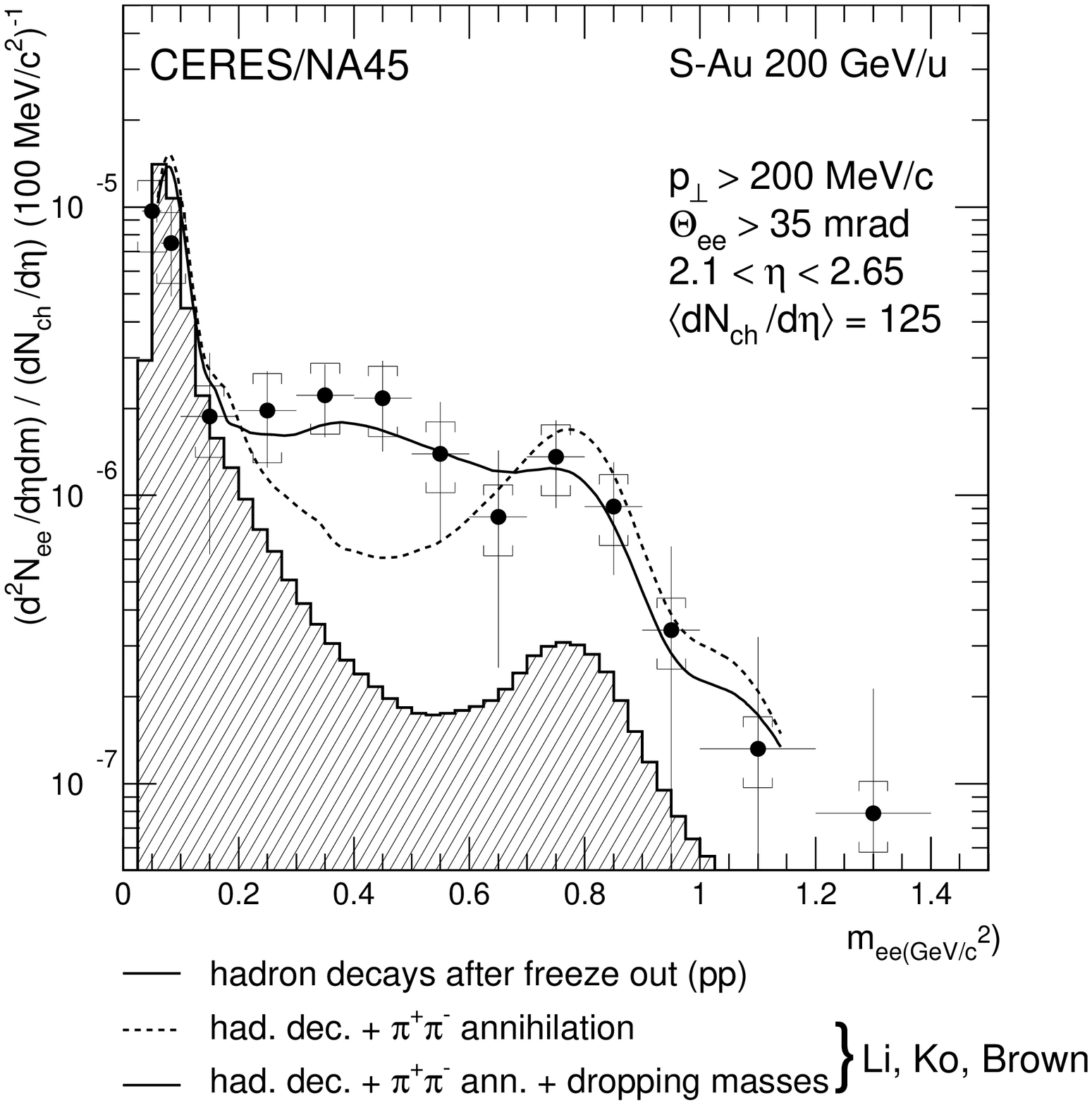}
\end{center}  
\vspace{-0.5cm}
\caption{Invariant mass spectrum of $e^+e^-$ measured by CERES in S-Au collisions
at 200 GeV/u compared to the expected yield from hadron decays (histogram) and to
the calculations of Li et al. \protect\cite{li-ko-brown} including in addition $\pi\pi$
annihilation (dashed line) and dropping $\rho$ mass (solid line).}
\end{figure}
As an example, we show in Fig. 8 one of the first calculations by Li, Ko and 
Brown \cite{li-ko-brown}.
 One sees that the addition of the $\pi^+\pi^-$ 
annihilation channel (dashed line in Fig. 8) on top of the hadronic sources 
listed above (histogram in Fig. 8), leads to a considerable increase of the 
low-mass electron pair yield particularly near the $\rho$ mass  --a direct 
consequence of the inclusion of the pion annihilation channel which is 
dominated by the pole of the pion form factor at the $\rho$  mass.  However, 
the calculation  fails to reproduce the data in the mass region  
0.2 $< m_{e^+e^-}<$0.5 GeV/c$^2$.

The same conclusion was reached by many other authors which have performed similar
calculations, including the pion annihilation channel, but treating the reaction
dynamics in completely different ways. Among those we quote
transport models which explicitly propagate baryons 
and mesons assuming the formation of a hadronic system in thermal equilibrium
\cite{li-ko-brown} or without  equilibrium \cite{cassing,koch,uqmd,haglin},
a standard hydrodynamical
\begin{figure}[h!]
\begin{center}
\leavevmode
\epsfxsize=12cm
\epsfysize=9cm
\epsffile{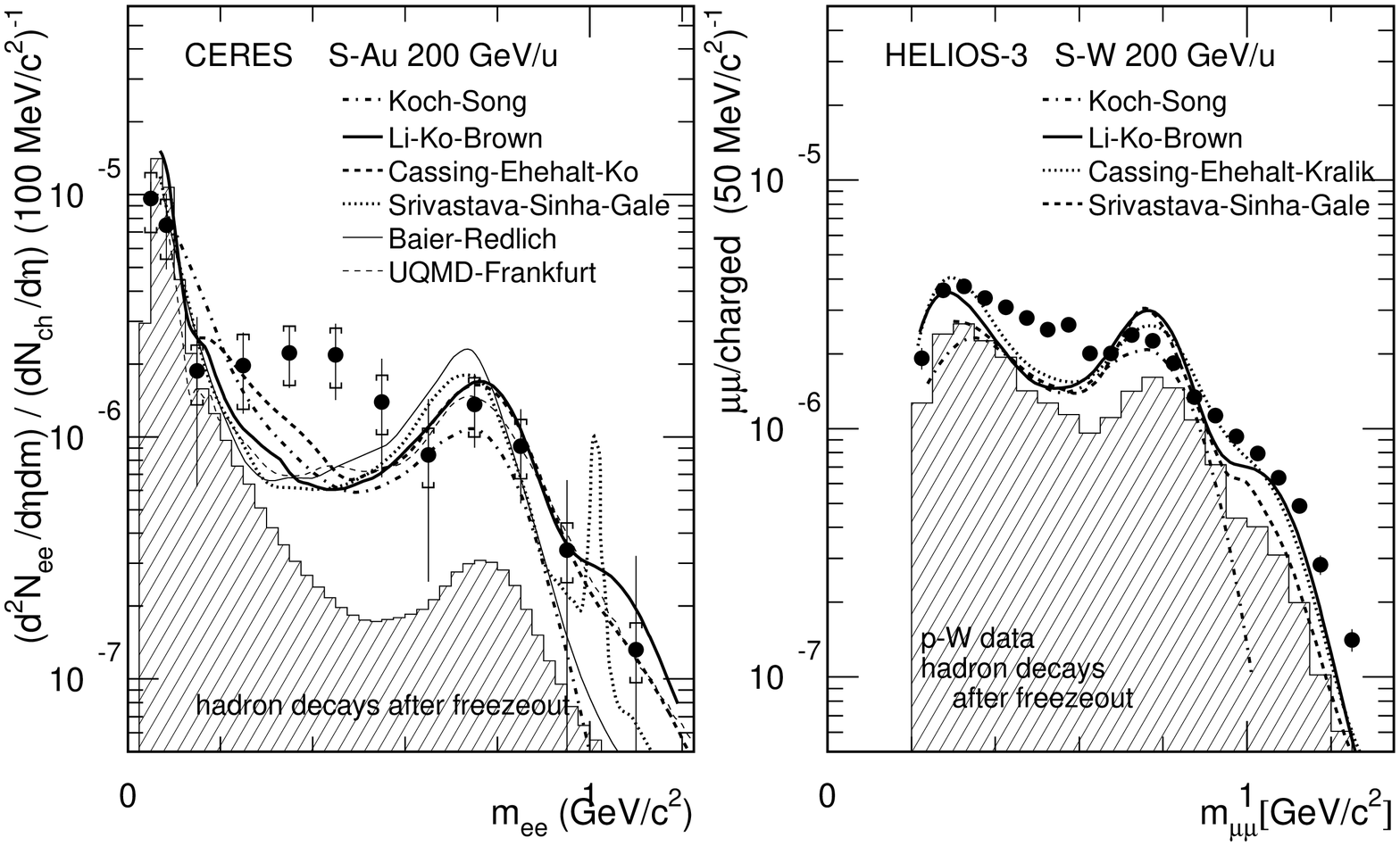}
\end{center}  
\caption{Comparison of the CERES and HELIOS-3 results with model 
calculations~\protect\cite{li-ko-brown,cassing}$^-$\protect\cite{srivastava} all 
including the two-pion annihilation channel.}
\end{figure}
model invoking or not invoking the formation of a thermalized QGP 
\cite{srivastava,shuryak,ruuskanen} 
and a model based on  a thermalized hadronic gas \cite{baier-redlich}. The common 
constraint is that the models are required to reproduce experimentally observed 
hadronic variables like  multiplicity, rapidity and p$_t$ distributions.
Some of those calculations are  shown in Fig. 9 in comparison with the CERES and 
HELIOS-3 results. A striking feature is that 
their predictions are very similar --within a factor of $\sim$2-- despite their 
different assumptions on  collision dynamics, indicating that the results are not 
very sensitive to the details of the space-time evolution of the collision. One sees, 
as in Fig. 8, that the pion channel accounts for a large fraction of the observed 
excess. The calculations reproduce well or even overshoot the dilepton yield near  
the $\rho$/$\omega$ mass. However, they all fail to reproduce the data at 
lower masses, in the region 0.2 $< m_{e^+e^-}<$0.5 MeV/c$^2$.

  Data in this mass region have been  quantitatively explained  
by taking into account the decrease of meson  masses --in particular of the $\rho$
meson-- in the hot and dense fireball as a precursor of chiral symmetry restoration 
\cite{pisarski82,brown91}.
\begin{figure}[h!]
\begin{center}
\leavevmode
\epsfxsize=12cm
\epsfysize=9cm
\epsffile{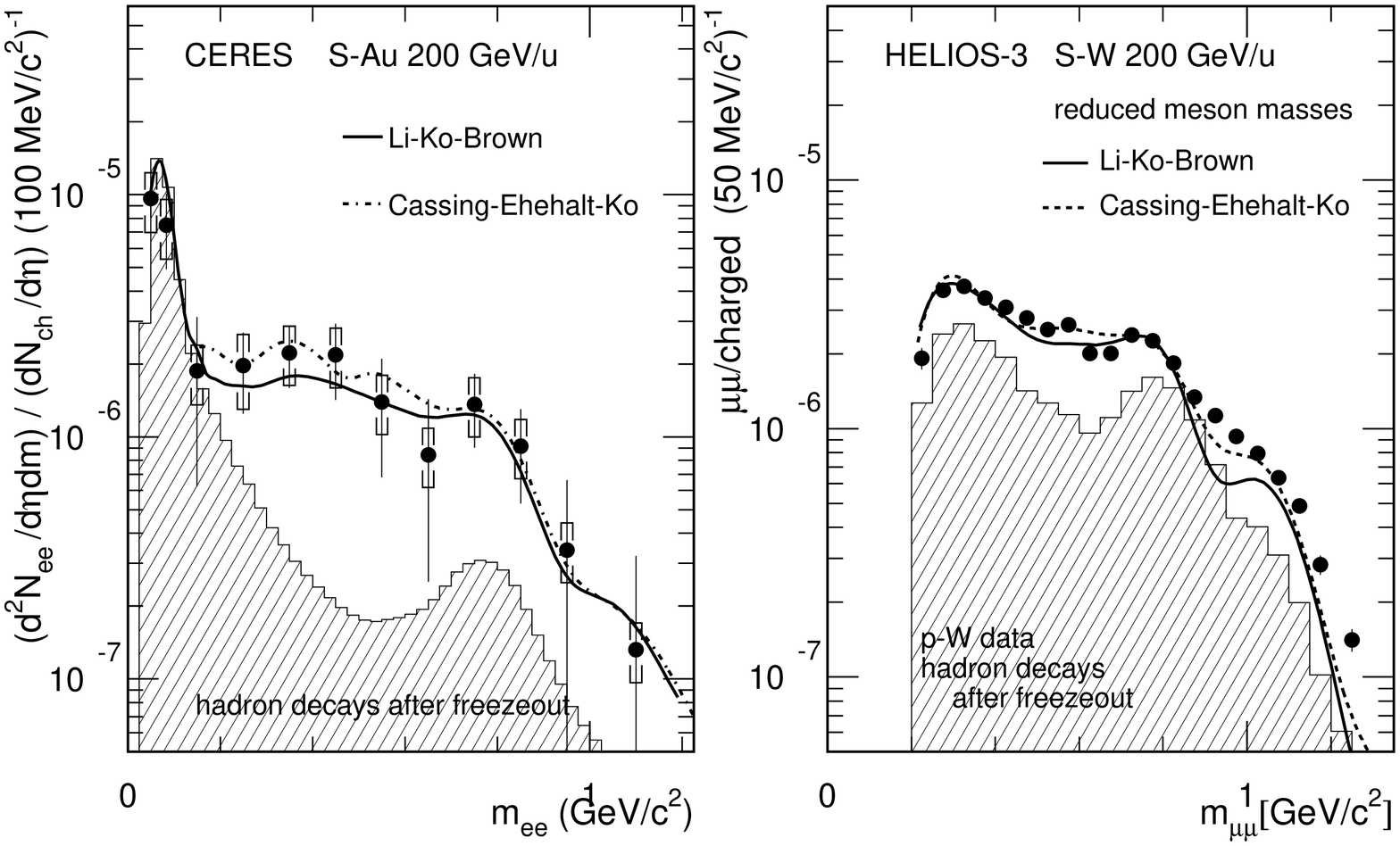}
\end{center} 
\caption{Comparison of the CERES and HELIOS-3 results with the calculations 
of refs.~\protect\cite{li-ko-brown,cassing}  with  dropping meson masses.}
\end{figure}
With this approach, first proposed by  Li, Ko and Brown \cite{li-ko-brown},
excellent agreement has been achieved with the CERES  
data, as shown in Fig. 8 by the solid line.  
Similar observations have been reported Cassing et al. \cite{cassing}.
They  derive the drop of the meson masses from QCD sum rules and their results 
are very similar to those of ref. 2. Both groups are also able 
to reproduce the HELIOS-3 low-mass enhancement. Their results are shown in Fig. 10.
The dropping mass scenario  also reproduces the enhancement observed in the preliminary
Pb-Au results from CERES \cite{ko-private}.

Other authors have investigated a different path, considering  modifications of the 
$\rho$-meson width in the dense medium, due to collision broadening. Although  a
larger width increases somewhat the yield of low-mass dileptons, the effect
is not strong enough to account for the oberved yield 
\cite{koch,haglin,wambach}. More recent calculations 
\cite{rapp-wambach,cassing-wambach}, have reached a different conclusion using a  
$\rho$-meson spectral function which  includes the pion modification in the 
nuclear medium and the scattering of $\rho$ mesons off baryons and mesons. This leads to  
a much larger broadening of the $\rho$-meson shape 
(see left panel of Fig. 11) and consequently to a considerable
enhancement of low-mass dileptons. These calculations are  able to 
reproduce very well the CERES and HELIOS-3  S data (see Fig. 11 right panel), and also the
CERES preliminary Pb-Au data, as well as the dropping $\rho$ 
mass scenario. These results are however not free of debate. Steele, Yamagushi and Zahed
\cite{zahed} addressed the same physics of in-medium modifications of the
$\rho$ spectral function.  Using on-shell chiral reduction
formulas and enforcing known constraints, they reached the conclusion that the additional
strength is not sufficient to explain the low-mass CERES data.

\begin{figure}[h!]
\begin{center}
\leavevmode
\hspace{-1.8cm}
\epsfysize=5.7cm
\epsffile{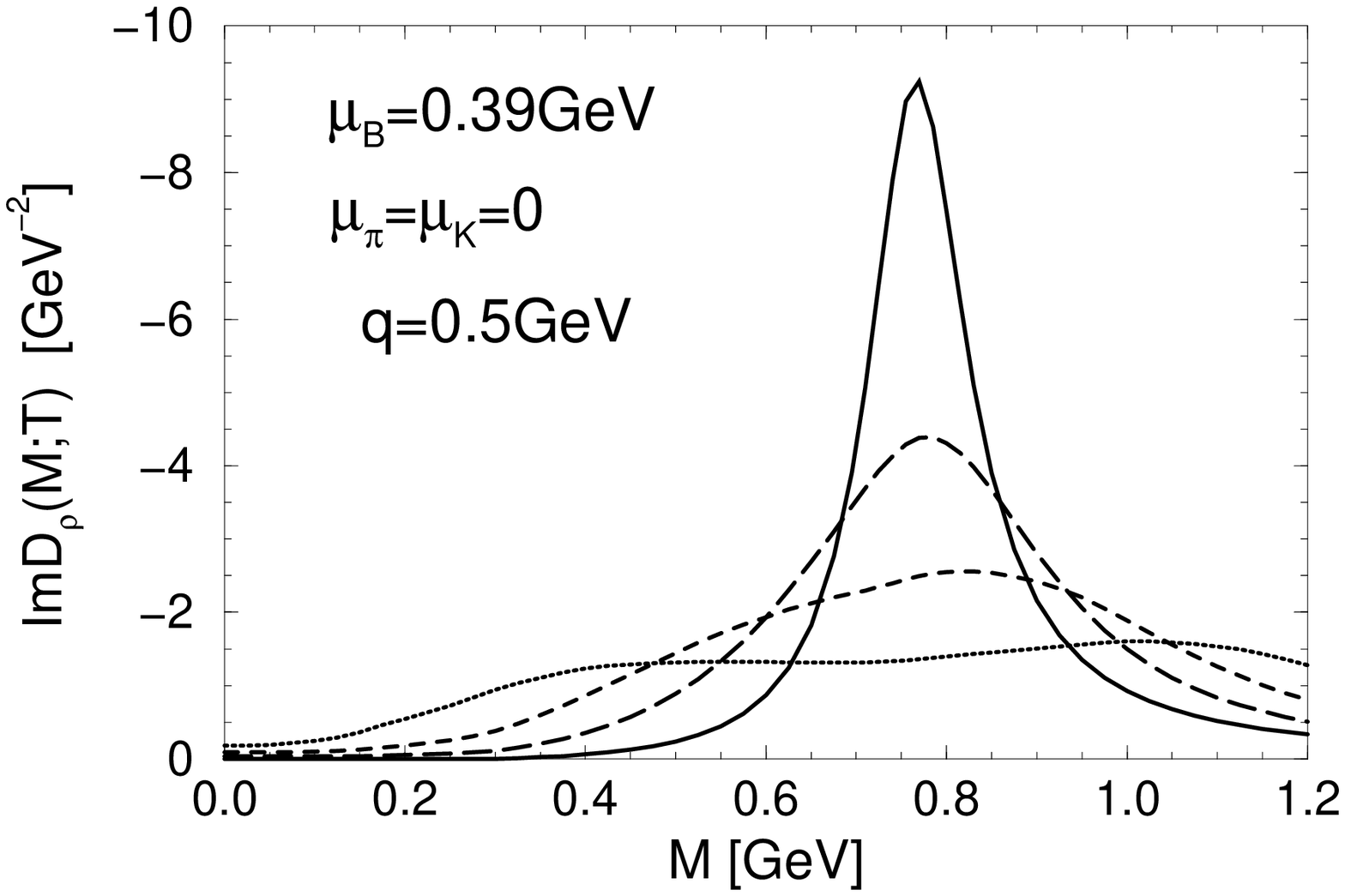}
\epsfysize=5.5cm
\epsffile{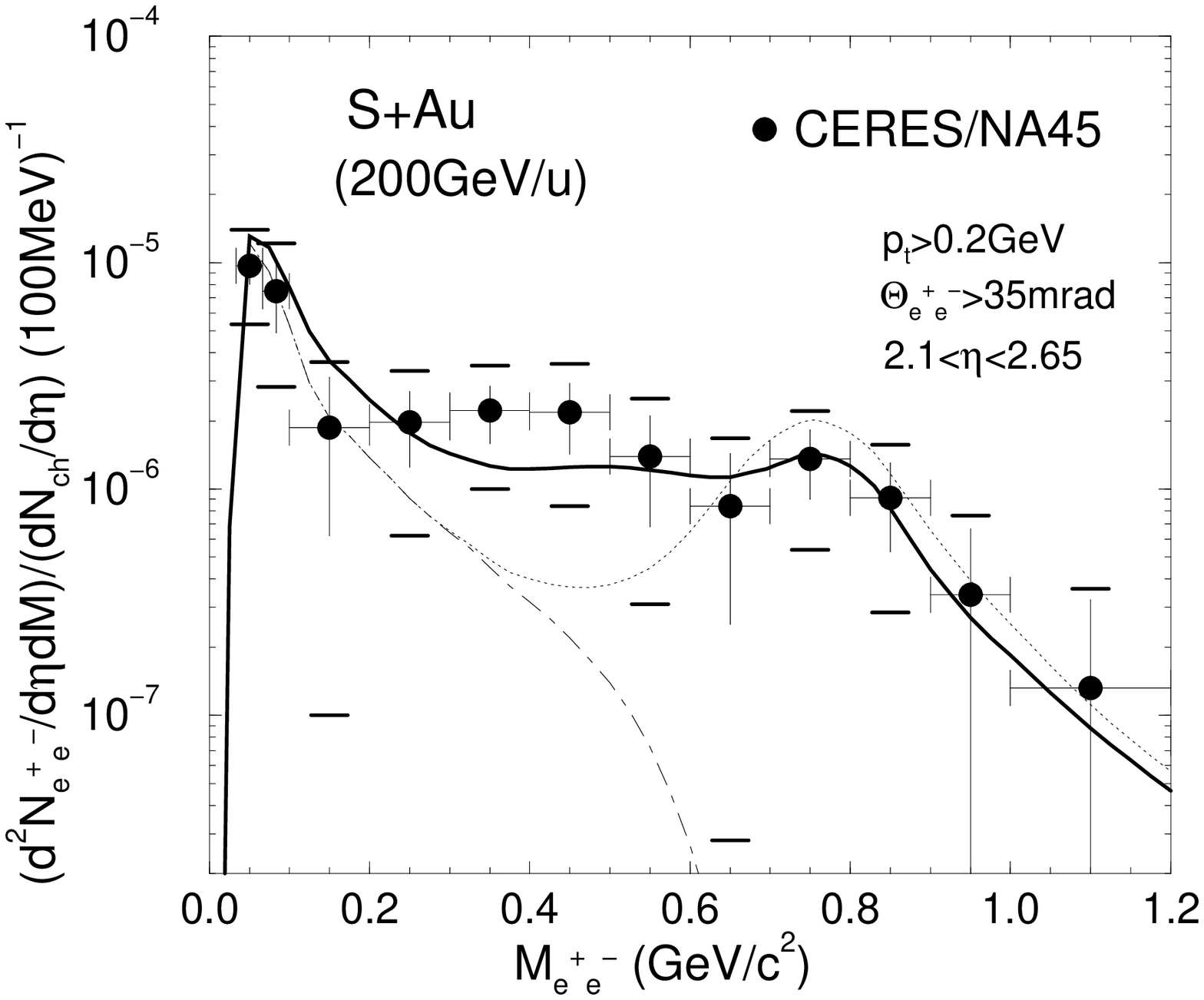}
\end{center}  
\caption{The  $\rho$-meson spectral function after inclusion of in-medium 
pion modifications and $\rho$ scattering off baryons and mesons. The curves correspond 
to fixed momentum (q=0.5 GeV) and chemical potentials ($\mu_{B}$=0.39 GeV, $\mu_{meson}$=0)
at temperatures of T=0.127 GeV (long-dashed), T=0.149 GeV (short-dashed) and
T=0.170 GeV (dotted) (left panel). 
Comparison of the resulting dielectron spectrum with the CERES data on S-Au collisions
\protect\cite{rapp-wambach} (right panel).}
\end{figure}

As discussed in the Introduction, direct photons should provide analogous information 
to thermal dileptons. Therefore, a simultaneous quantitative  description of results 
on low-mass dileptons and direct photons  within a single model would be very 
decisive in establishing a consistent and reliable interpretation 
of experimental results. In this context, the lack of signal in the photon data is in 
striking contrast to  the strong enhancement of low-mass dileptons and raises the 
question of consistency between the two experimental findings.
In fact the real question is a quantitative one, namely the level of 
sensitivity of the two measurements to a new source with respect to their hadronic 
background. Simple arguments reveal that the level of sensitivity is more than two 
orders of magnitude lower for photons compared to electrons~\cite{it-qm95}. In other 
words, the enhancement factor of 5 observed in the CERES S-Au data should translate 
into an enhancement factor of a few percent in the photon measurement.
This conclusion has been recently confirmed by calculations performed by Li and Brown
\cite{li-photons} with the same model of dropping meson masses  used to explain the 
CERES and HELIOS-3 low-mass dilepton results.
\begin{figure}[h!]
\begin{center}
\leavevmode
\epsfxsize=9cm
\epsfysize=8cm
\epsffile{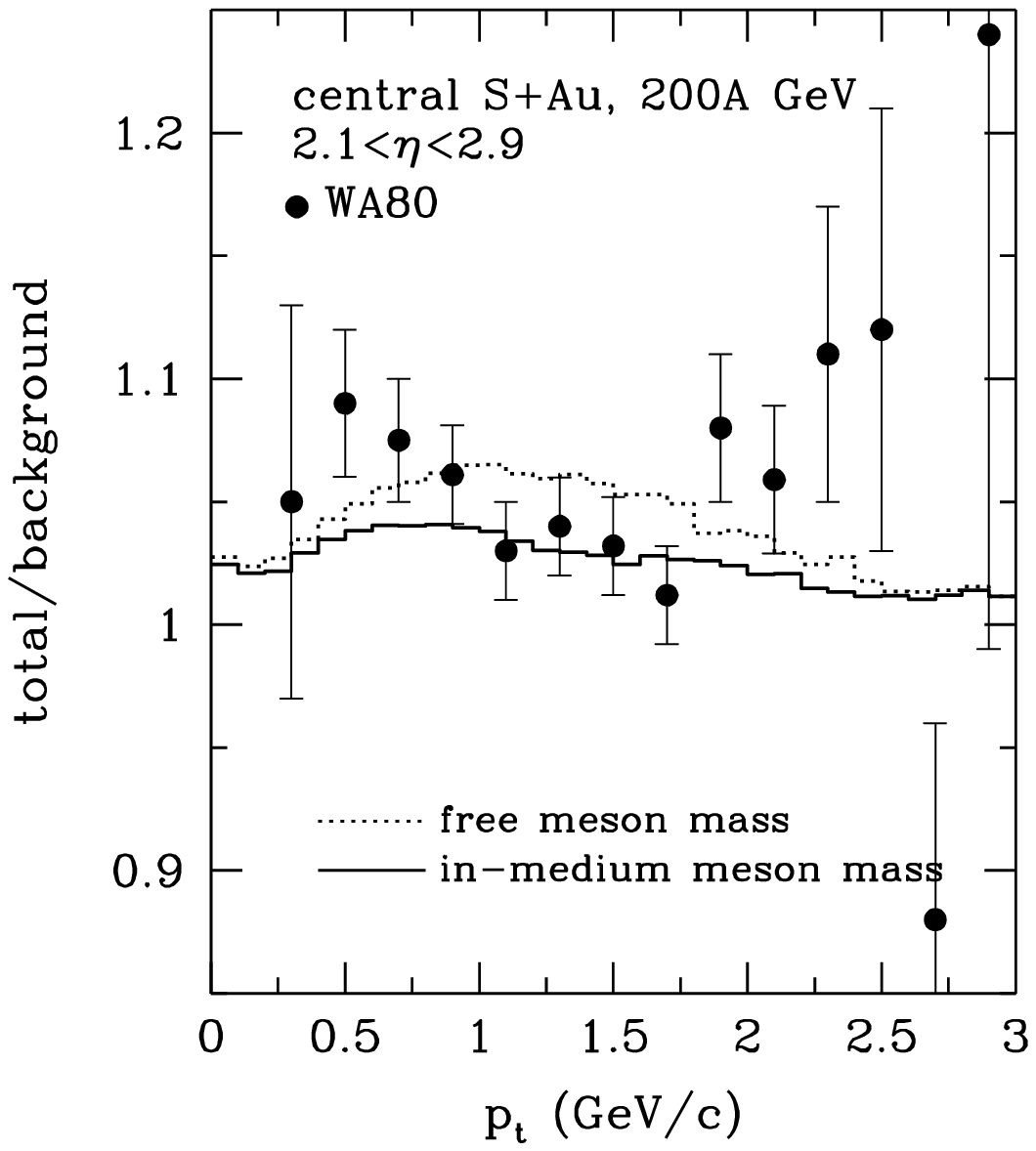}
\end{center} 
\caption{Comparison of the WA80 photon data (presented as the ratio of the total 
measured yield to the expected background yield from $\pi^0$ and $\eta$ decays) with the 
calculations 
of refs.~\protect\cite{li-photons} with (solid line) and without (dotted line)
dropping meson masses.}
\end{figure}
 Their results are shown in Fig. 12 and compared to the WA80 data. The excess of 
direct photons is predicted to be a few
percent of the total hadronic background, in agreement with the experimental results.
This sharpens the strict requirement imposed on the experiments to control the 
systematic errors down to the percent level in order to be able to observe direct photons.

We finally turn  briefly to the intermediate mass region.
An interesting question is whether the excess in this mass region 
has the same origin as at low-masses. In the HELIOS-3 data the 
mass spectrum of the excess has the same slope below and above the vector mesons 
$\rho$, $\omega$ and $\phi$, suggesting a common origin \cite{masera-qm95}. NA38 
however, makes the observation that the shape of the excess in the intermediate 
mass region resembles very much the shape of the open charm contribution suggesting 
that enhanced charmed production could be at the origin of the excess.  
One would then need a different explanation for the excess at low masses since 
charm production has a negligible contribution there. Data on multiplicity
dependence and p$_t$ distribution may shed light on this issue.

\section{Summary and Outlook} 
  
CERES, HELIOS-3 and NA38 have observed a significant excess of lepton
pairs in S-Au collisions at 200 GeV/nucleon over the expected yield of
known hadronic sources. 
The excess is confirmed by  results obtained by  the CERES 
and  NA50 experiments in 160 GeV/u Pb-Au and Pb-Pb collisions, respectively.
Theoretical calculations show that a large fraction of the low-mass 
(m = 0.2 - 1.5 GeV/c$^2$) excess  originates from the pion annihilation into
lepton pairs, $\pi^+\pi^- \rightarrow \rho \rightarrow l^+l^- $, thereby 
providing the first evidence  of thermal radiation emitted from the dense 
hadronic matter formed in these collisions. The models achieve an excellent 
agreement with all data sets by further requiring  strong in-medium modifications
of the vector mesons, in particular a decrease of the $\rho$-meson mass as a
precursor of chiral symmetry restoration.

 There is no evidence of direct photon emission in these collisions and  
the various experiments allow to set an upper limit of the 
order of 10\%. The lack of signal is attributed to the lower sensitivity
of the photon measurement as compared to dileptons.  The same model of dropping 
meson masses provides a good description of the photon data.

  The dilepton experimental results together with the hints of in-medium effects 
and  chiral symmetry restoration have created a considerable excitement. 
In order to further constrain the models we need more and precise information 
on the excess: the two key questions are the p$_t$ distribution and the 
multiplicity dependence of the observed excess. A considerable progress is 
expected from the present round of experiments at the CERN SPS with the Pb beam
and detailed information on multiplicity and p$_t$ dependences should be available
within a year.
Two major new steps are foreseen in a somewhat longer time scale. First, 
 CERES is planning to dramatically improve the
mass resolution to achieve $\delta m/m $= 1\%,  by the addition of a TPC 
downstream of the present double RICH spectrometer. With this resolution, 
which is of the order of the natural line width of the $\omega$ meson,
it should be possible to  directly measure the yield of all three vector 
mesons $\rho, \omega$ and $\phi$ including any possible changes in their 
properties (mass shift or increased width) thereby providing compelling
evidence of the scenarios invoking chiral symmetry restoration. Second,
a measurement is proposed at the lowest energy attainable at the SPS, at about
40 GeV/nucleon, where the effect of baryon density on the vector meson masses 
is expected to be largest.

\section*{Acknowledgments}
 This work was supported by the Israeli Science Foundation, the MINERVA Foundation,
the German-Israeli Science Foundation for Scientific Research and Development, and 
the Benoziyo High Energy Research Center.

\end{document}